\documentclass[twocolumn, prd,preprintnumbers,superscriptaddress,nofootinbib,showpacs]{revtex4-1}

\usepackage{amsfonts,amssymb,amsmath}
\usepackage{mathrsfs}
\usepackage{color}
\usepackage{graphicx}
\usepackage{dcolumn}
\usepackage{soul}

\usepackage{bm}

\def\lsim{\mathrel{\rlap{\lower4pt\hbox{\hskip1pt$\sim$}}
    \raise1pt\hbox{$<$}}}         %less than or approx. symbol
\def\gsim{\mathrel{\rlap{\lower4pt\hbox{\hskip1pt$\sim$}}
    \raise1pt\hbox{$>$}}}         %greater than or approx. symbol

\begin{document}

\preprint{APCTP Pre2017-011}

\title{Clockwork seesaw mechanisms}

\author{Seong Chan Park}
\email{sc.park@yonsei.ac.kr}
\affiliation{Department of Physics and IPAP, Yonsei University, Seoul 03722, Korea}
\author{Chang Sub Shin}
\email{changsub.shin@apctp.org}
\affiliation{Asia Pacific Center for Theoretical Physics, Pohang 37673, Korea\\
Department of Physics, Postech, Pohang 37673, Korea}
\affiliation{Department of Physics, Postech, Pohang 37673, Korea}

\date{\today}

\begin{abstract}
We propose new mechanisms for small neutrino masses based on clockwork mechanism. %Majorana mass of neutrino.  
The Standard Model neutrinos and  lepton number violating operators communicate %with each other 
through the zero mode of clockwork gears, one of the two couplings of the zero mode is 
exponentially suppressed by clockwork mechanism.   Including all known 
examples for the clockwork realization of the neutrino masses,  different types of models are realized depending on the profile and chirality of the zero mode fermion. Each type of realization would have phenomenologically distinctive features with the accompanying heavy neutrinos. 

\end{abstract}

\pacs{
14.60.Pq,	%Neutrino mass and mixing (see also 12.15.Ff Quark and lepton masses and mixing)
14.60.St	%Non-standard-model neutrinos, right-handed neutrinos, etc.
%14.80.Bn %Standard-model Higgs bosons
%14.65.Ha	%Top quarks
%95.35.+d	%Dark matter (stellar, interstellar, galactic, and cosmological) 
%97.60.Lf	%Black holes
%98.80.Cq	%Particle-theory and field-theory models of the early Universe (including cosmic pancakes, cosmic strings, chaotic phenomena, inflationary universe, etc.)
}
\keywords{neutrino mass, clockwork mechanism, Majorana mass}

\maketitle

%%%%%%%%%%%%%%%%%%%%%%%%%%%%%%%%%%%%%%%%%%%%%%%%
\section{introduction}
\label{sec:introduction}
%%%%%%%%%%%%%%%%%%%%%%%%%%%%%%%%%%%%%%%%%%%%%%%%

Generating hierarchies among couplings or scales have been regarded as important problems in particle physics in various  contexts.  One of such hierarchies is related with neutrino masses:  whose values are unknown so far but bounded from above $m_\nu \lsim 0.1$ eV~\cite{KamLAND-Zen:2016pfg}, which is enormously small compared to any mass scale of charged fermion. 

%seesaw
The most popular explanation of the hierarchy is the seesaw mechanism \cite{Minkowski:1977sc,Yanagida:1979as, GellMann:1980vs, Mohapatra:1979ia}.  The large hierarchy between  the weak scale and  the Majorana mass of the additional SM singlet fermion, $m_M$, is converted into the hierarchically small neutrino mass,  $m_\nu\sim (yv)^2/m_M$ with the Yukawa coupling $y$, which is presumably of the order of unity from our sense of naturalness, and the vacuum expectation value of the Higgs field $v=246$ GeV. On the other hand, the hierarchy $v/m_M \ll 1$ could be destabilized  due to the large quantum corrections to the Higgs mass square, $\delta m_H^2 \sim y^2 m_M^2$,  thus a  large Majorana mass, $m_M$, is not desirable after all. Instead, if $m_M$ is chosen to be around weak scale to avoid the large radiative correction, a small Yukawa coupling, $y\lsim 10^{-6}$,  is requested for small neutrino mass, which calls for additional model building.
%inverse seesaw
An interesting variation of seesaw mechanism is inverse seesaw mechanism.  By introducing one more single fermion with  Dirac ($M_D$) and Majorana ($m_M$) mass terms, the neutrino mass is suppressed not by $1/m_M$ but by $m_M$:
$m_\nu \sim m_M \frac{ (yv)^2}{M_D^2 + (yv)^2}$. If we assume the weak scale Dirac mass parameters, $M_D\sim  v$, in order not to destabilize the hierarchy between the weak scale and the  Dirac mass scale,  a small neutrino mass $m_\nu \sim m_M$ is realized only by a small Majorana mass.  This setup is technically natural as the vanishing Majorana mass is consistent with the lepton number conservation \cite{tHooft:1979rat}. However, the small value of Majorana mass itself calls for additional explanation again (see e.g. \cite{Park:2009cm} for extra dimensional explanation (also see \cite{Agashe:2016ttz, Agashe:2017ann})).

%%%%%%%%%%%%%%%%%%%%%%%%%%%%%%%%%%%%
\begin{figure}[t]
\begin{center}
\includegraphics[width=0.47\textwidth]{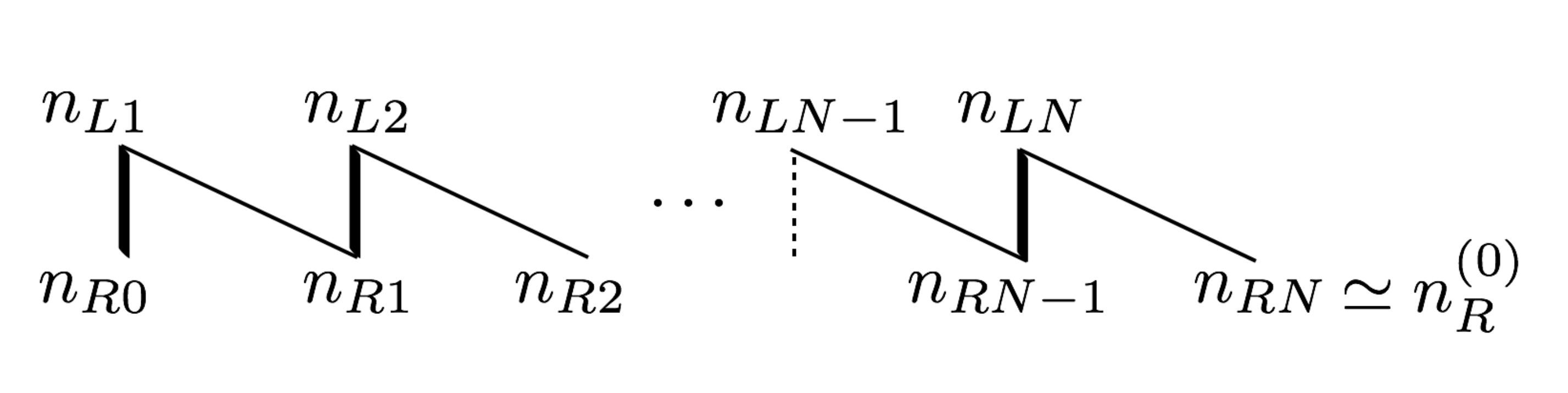} %figures directory 
\includegraphics[width=0.47\textwidth]{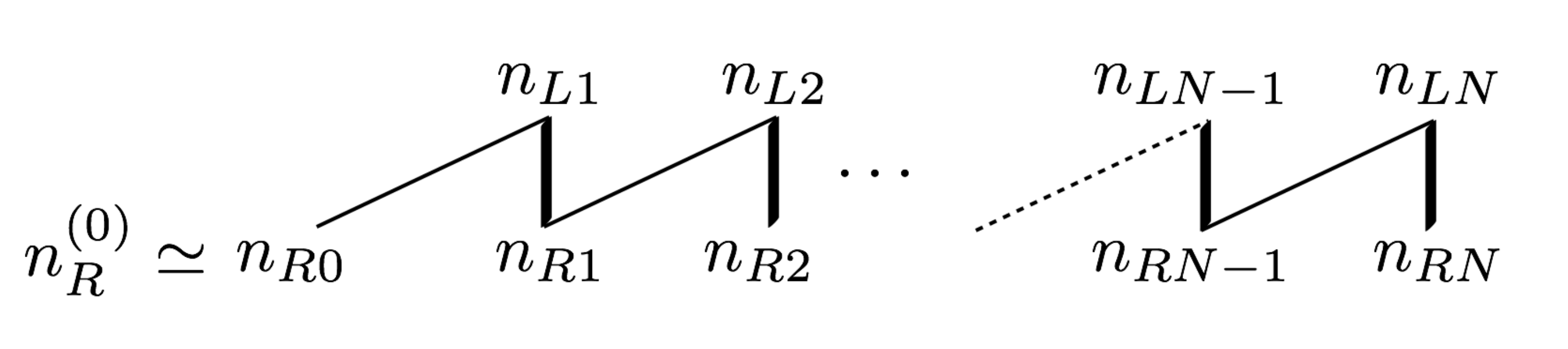} %figures directory 
\end{center}
\caption{The CW diagrams for $V_{{\rm CW}_{R,R}}$ and $V_{{\rm CW}_{R,L}}$ in Eq.~(\ref{clockwork_potR}) and Eq.~(\ref{clockwork_potL}), respectively provides the right-chiral zero mode field, $n_{R}^{(0)}$, localized at $i=0$ and  $i=N$.  One can similarly construct the potentials $V_{{\rm CW}_{L,R}}$, $V_{{\rm CW}_{L,L}}$ and diagrams with a left-chiral zero mode localized at $i=0$ and $i=N$, respectively.  The extra dimensional interpretation is obvious: the localization of a chiral field at left or right boundary with Kaluza-Klein modes represented by CW gears.}
\label{fig:CW diagram}
\end{figure}
%%%%%%%%%%%%%%%%%%%%%%%%%%%%%%%%%%%%

In this letter, we propose new simple ways to generate such small Yukawa coupling or Majorana mass using the {\it clockwork(CW) mechanism} taking the advantage of generating hierarchical structure of zero mode (massless mode) couplings to matter without introducing any unnaturally small couplings and/or scales by {\it the CW gears} (see Fig.~\ref{fig:CW diagram}).  The main idea of clockwork mechanism was first proposed in order to  generate a tans-plankian period of the pseudo scalar inflaton potential~\cite{Choi:2014rja}, and utilized  in more general cases \cite{Choi:2015fiu,Higaki:2015jag,Kaplan:2015fuy}.  It is also generalized to the fields with different spins, and recognized that the localization of the wave functions in the site space resembles that in the deconstruction of the extra dimensional model \cite{Giudice:2016yja}, although the details are not exactly the same \cite{Craig:2017cda,Giudice:2017suc}. 
There are also interesting applications of the mechanism for various phenomenological problems such as dark matter, inflation, composite Higgs, axion and others~\cite{Kehagias:2016kzt,Farina:2016tgd,Ahmed:2016viu,Hambye:2016qkf,Coy:2017yex,Ben-Dayan:2017rvr}.    We also note that small neutrino masses are discussed in CW mechanism with and without Majorana mass term in \cite{Hambye:2016qkf} and \cite{Giudice:2016yja}, respectively. (see Sec.~\ref{c_s} for details.)

In the next section(Sec.~\ref{c_g}) we first review the basic idea of clockwork mechanism for our purpose  then apply to the seesaw models in the following sections (Sec.~\ref{c_s} and Sec.~\ref{c_i}), where we also discuss the extra dimensional interpretation of CW models. Finally we discuss distinctive phenomenological features of different realizations then conclude in the last section (Sec.~\ref{sec:discussion}).

\section{clockwork mechanism} \label{c_g}

In CW theories, a (large) number of fields, called CW gears, are introduced, linked and generate an amplified structure just as a series of gears in a machine can generate a large movement of the last gear by a small movement of the first gear.
A CW theory may be interpreted as a deconstructed version of an extra dimensional model as we will show more details later.

For our purpose here, we consider a set of left handed $(n_{Li})$ and right handed $(n_{Ri})$ fermions. 
Depending on the chirality of zero mode state, we actually have two options as 
\begin{eqnarray}
{\rm CW}_R&:&\, 
n_{Li}\ (i=1,\cdots, N) \nonumber\\
&&\, n_{Ri}\ (i=0,1,\cdots, N), \\
{\rm CW}_L&:&\, 
n_{Li}\ (i=0,1,\cdots, N) \nonumber\\
&&\, n_{Ri}\ (i=1,\cdots, N), 
\end{eqnarray}
where the unpaired degree of freedom i.e. $n_{R0}$ in ${\rm CW}_R$ or $n_{L0}$ in ${\rm CW}_L$ is for the zero mode or the lightest physical degree of freedom in particle spectrum. For ${\rm CW}_R$, two CW potentials can be constructed assuming slight hierarchy in the mass parameters, $M> m$ as 
\begin{eqnarray}
\label{clockwork_potR}
V_{{\rm CW}_{R,R}}\equiv\sum_{i=1}^{N} M\, \bar n_{Li}  n_{R\, i-1} - m\, \bar n_{Li} n_{R i}, \\
\label{clockwork_potL}
V_{{\rm CW}_{R,L}}\equiv\sum_{i=1}^{N}  M\, \bar n_{Li} n_{R i}-m\, \bar n_{Li}  n_{R\, i-1} . 
\end{eqnarray}
$M$ and $m$ could be considered as spurion fields that break  chiral symmetries of the gear fermions. 
The corresponding CW diagrams are depicted in Fig.~\ref{fig:CW diagram} where the thick lines represent the mass terms of $M$ and thin lines the terms of $m$, respectively. Similarly, one can also get a mass potential for ${\rm CW}_{L}$, $V_{{\rm CW}_{L, R}}$ and $V_{{\rm CW}_{L, L}}$, by changing the role of $n_L$ and $n_R$, respectively. 

In order to  get an effective action for the light fields, one can use the Euler-Lagrangian equations of motion for the heavy fields $  n_{Li}$. For $V_{{\rm CW}_{R,R}}$, as an explicit example,  we get
\begin{eqnarray}
M\, n_{R\, i-1} = m\, n_{R i}, \ (i=1,2,\cdots, N),
\end{eqnarray}
or
\begin{eqnarray}
n_{R0} &=& \frac{m}{M}n_{R1} =\left(\frac{m}{M}\right)^2 n_{R2} \nonumber \\
\cdots&=& \left(\frac{m}{M}\right)^i n_{Ri}\cdots=\left(\frac{m}{M}\right)^N n_{RN},
\end{eqnarray}
where $0\leq i \leq N$. 
Then,  it is easily identified that the clockwork zero mode, $n_R^{(0)}$,  is quasi-localized at the $(N+1)$th CW gear ($i=N$),
\begin{eqnarray}\label{CWzeromode}
 n_{Ri}  \simeq \left(\frac{m}{M} \right)^{N-i} n_R^{(0)}~~~\text{(for $V_{{\rm CW}_{R,R}}$)}.
\end{eqnarray}
Although the equations of motion are evaluated for $m\ll M$,  Eq.~(\ref{CWzeromode}) is still valid as long as  $m\lesssim M$, because it is related with the constant shift symmetry for the CW zero mode:
$n_R^{(0)}\to n_R^{(0)} + \alpha_R$. 
Given the unitary transformation of the gear fields to the CW mass eigenstates, 
$n_{Ri} = \sum_{n=0}^NC_{i (n)} n_{R}^{(n)}$, the shift symmetry of the zero mode corresponds to the transformation of $n_{Ri}$ as $n_{Ri} \to n_{Ri} +  C_{i(0)}\alpha_R$.  The potential, $V_{{\rm CW}_{R,R}}$ should be invariant under the transformation, so that the mixing coefficients   satisfy $M C_{i-1(0)} -m C_{i (0)}=0$ for $i=1,\cdots, N$. 
After integrating out massive states (equivalently taking $n_R^{(n)}=0$ for $n=1,\,\cdots, N$), we get $n_{Ri} = C_{N(0)}(m/M)^{N-i} n_R^{(0)}$, where the normalization factor $C_{N(0)}$ is ${\cal O}(1)$ for $m\lesssim M$, so we safely ignore it for the simplification. 

The role of $M$ and $m$ is changed for $V_{{\rm CW}_{R,L}}$,  the zero mode is localized at the first ($i=0$) CW gear as
\begin{eqnarray}
 n_{Ri}  \simeq \left(\frac{m}{M} \right)^{i} n_R^{(0)}~~~\text{(for $V_{{\rm CW}_{R,L}}$)}.
\end{eqnarray}
By construction, a small hierarchy $m<M$ would induce a large suppression factor $(m/M)^N\ll 1$, which may provide a natural explanation of small parameters (couplings or mass scales) in phenomenological models:
This is the most notable merit of CW mechanism.
%explain small couplings or small scales, which would look apparently unnatural. This is the essential idea of CW mechanism. 

In the following sections, we will apply this attractive features of CW mechanism to understand the small neutrino masses.  In our explicit model construction, we can take one of the four potentials $V_{{\rm CW}_{R,R}}$, $V_{{\rm CW}_{R,L}}$, $V_{{\rm CW}_{L,R}}$ and $V_{{\rm CW}_{L,L}}$\footnote{Mathematically,  all these constructions are equivalent, when we relabel the states by $L\to R$, and $i$ to $N-i$. However, in this letter, we distinct each cases by imposing the lepton numbers  as $n_{ l}(n_{Li}) = - n_{  l}(n_{Ri})= 1$ and 
taking the state with $i=0$ as the only gear field that couples to the SM sector. }, which lead to different chiral zero modes and different localization sites then different phenomenological features, respectively.

%%%%%%%%%%%%%%%
\section{Type I Clockwork seesaw}\label{c_s}
%%%%%%%%%%%%%%%

As it is well known, an extension of the standard model with a singlet right chiral fermion, $n_R$, would allow Yukawa interaction and Dirac mass for a neutrino: $y  \bar\ell_L H n_R \to y v \bar\nu_L n_R$ where the extremely small Yukawa coupling constant is requested as $y = m_\nu/v \sim 10^{-12}$. A simple explanation is discussed in \cite{Giudice:2016yja} in the context of CW mechanism just by adding the Yukawa interaction between the lepton doublet and $n_{R0}$, $y \bar l_L H  n_{R0} + h.c.$, to the CW mass terms:
\begin{eqnarray}
V_0=   y v \bar\nu_L n_{R0}  + V_{{\rm CW}_{R,R}} + h.c.
\end{eqnarray} 
For $y v$, sufficiently smaller than $M$,   after integrating out all heavy modes, we get 
$n_{R0} \simeq (m/M)^{N} n_R^{(0)}$, and 
the effective 
potential for the light fields as 
\begin{eqnarray}
V_{0, \rm eff} \simeq  y_{\rm eff} v \bar \nu_L n_R^{(0)} + h.c.,
\end{eqnarray}
which implies 
\begin{eqnarray}
m_\nu \simeq y_{\rm eff} v,
\end{eqnarray}
where $y_{\rm eff} = y (m/M)^{N}$.
With $m/M \sim 1/3$, for instance, and $N=25$, the effective Yukawa coupling is quite small as $y_{\rm eff} \sim 10^{-12}$ so that the hierarchy of neutrino mass ($m_\nu \sim 0.1$ eV) and the electroweak scale ($v\sim 10^{11}$ eV) is understood even if the original Yukawa coupling is $y \sim 1$.

\begin{figure}[t]
\begin{center}
\includegraphics[width=0.43\textwidth]{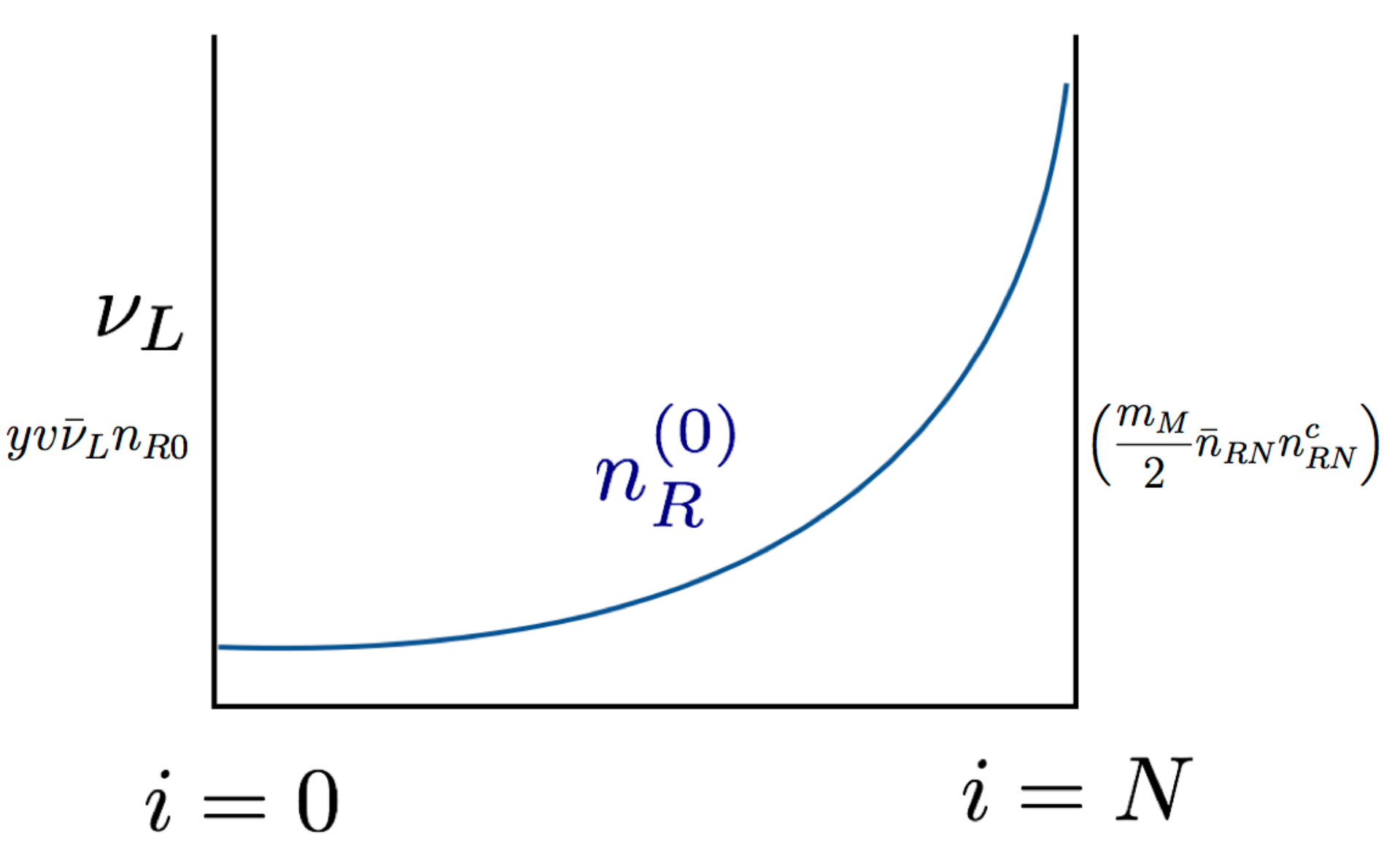} %figures directory 
\end{center}
\caption{Type I CW seesaw with the potential $V_0$ or $V_0+(\delta V_0)$.}\label{Dirac}
\end{figure}

If we break the lepton number at the end of CW gears by introducing the Majorana mass term  for $n_{R N}$,  
\begin{eqnarray}
\delta V_0 = \frac{m_M}{2} \bar n_{R  N} n_{R N}^c + h.c., 
\label{eq:maj}
\end{eqnarray}
the SM neutrino now gets a Majorana mass \cite{Hambye:2016qkf}:
\begin{eqnarray}
m_\nu \simeq \frac{y_{\rm eff}^2 v^2}{m_M}~~\text{(Type I)}.
\label{eq:double}
\end{eqnarray}
This is the usual   Type I seesaw mechanism in the sense that  the source of the lepton number violation is provided by the SM singlet fields, and vanishing $m_M$ just gives the Dirac mass. However the effective Yukawa interaction itself is  doubly suppression as $y_{\rm eff}^2 \propto (m/M)^{2N}$, so that the necessary number of $N$ is reduced compared to the case without the Majorana mass term and  there is no need for a large $m_M$ differently from the original seesaw mechanism. Fig.~\ref{Dirac} shows the main idea in this section: at $i=0$ site, the Yukawa term is introduced and the CW gears provide a link between the end points $i=0$ and $i=N$. At $i=N$ site, the would-be-zero-mode $n_{RN}\simeq n_R^{(0)}$ is localized and its Majorana mass term is introduced. %to get the double suppression factor in Eq.~\ref{eq:double}.  

%%%%%%%%%%%%%%%%%%%%%%%%%%%%%%%%%%%%
%%%%%%%%%%%%%%%%%%%%%%%%%%%%%%%%%%%%

It is intriguing to note that the constructed model has a natural extra dimensional interpretation: the Dirac mass term is localized on one boundary of extra dimension (say, at $x_5=0$) and the Majorana mass term is on the other side (say, at $x_5= L$ with a compactification length $L$).  The bulk field provides communications between the interactions localized at the opposite boundaries (see e.g. \cite{Grossman:1999ra}).

%%%%%%%%%%%%%%
\section{ClockWork inverse seesaw} \label{c_i}
%%%%%%%%%%%%%%

In a sense type I CW seesaw relies almost entirely on  the suppression in the effective Yukawa coupling even the precise structure of the mass matrix would differ depending on the presence of Majorana mass term. 
Here we propose different realization of CW seesaw where the role of the non-vanishing Majorana mass is essential. 
As it will be clear later on, our setup eventually has similarities with the inverse seesaw models but the CW mechanism would provide natural realization of small parameters, which are compulsory in successful inverse seesaw models. 

In conventional inverse seesaw models, three neutrinos (one SM doublet: $\nu_L$, two SM singlets: $n_R,\, N_L$) are introduced with a mass potential, 
\begin{eqnarray}
V =  y v \overline{\nu}_L n_R + M_D \bar N_L n_R + \frac{m_M}{2} \bar N_L N_L^c+h.c.,
\end{eqnarray}
where $M_D$ and $m_M$ are Dirac and Majorana mass parameters, respectively. 
For small values of Yukawa coupling or Majorana mass term, the neutrino mass has the form as 
\begin{eqnarray}
m_\nu\simeq m_M   \frac{y^2 v^2}{M_D^2 +y^2 v^2}
\label{eq:mass IS}
\end{eqnarray}
as  discussed in the introduction. 
It is obvious from the expression that the correct neutrino mass scale, $m_\nu\sim 0.1$ eV, is realized in two parameter choices: 
\begin{eqnarray}
&&\textrm{Case a}: \  y v \ll  m_M \sim M_D,  \nonumber\\
&&\textrm{Case b}:\  m_M \ll y v \sim M_D,
\end{eqnarray}
however neither case would  have a natural explanation by itself. We will show how the CW mechanism would support these cases  without tuning parameters. 

Before going to the specific realization, we would mention some advantage  of CW realization of the inverse seesaw. First of all, one of the most interesting features of the inverse seesaw mechanism is the presence of relatively light new states with sizable couplings to the SM particles, which may show up at collider experiments, provided that the Majorana mass term is  small.  Those new states have   specific pattern of the mass spectrum  including CW gear fields by construction as discussed in Sec.~\ref{sec:discussion}.  We can  understand the origin of neutrino mass not only just from clockwork mechanism but also from specific realization of seesaw mechanisms by studying their collider phenomenology.
Secondly, the role of the lepton number violating interactions to the neutrino mass is more evident in the CW inverse seesaw realization compared to the Type I seesaw model.

%%%%%%%%%%%%%%%%%%%%%%%%%%
\subsection{Type Ia CW seesaw}
%%%%%%%%%%%%%%%%%%%%%%%%%%

\begin{figure}[t]
\begin{center}
\includegraphics[width=0.464\textwidth]{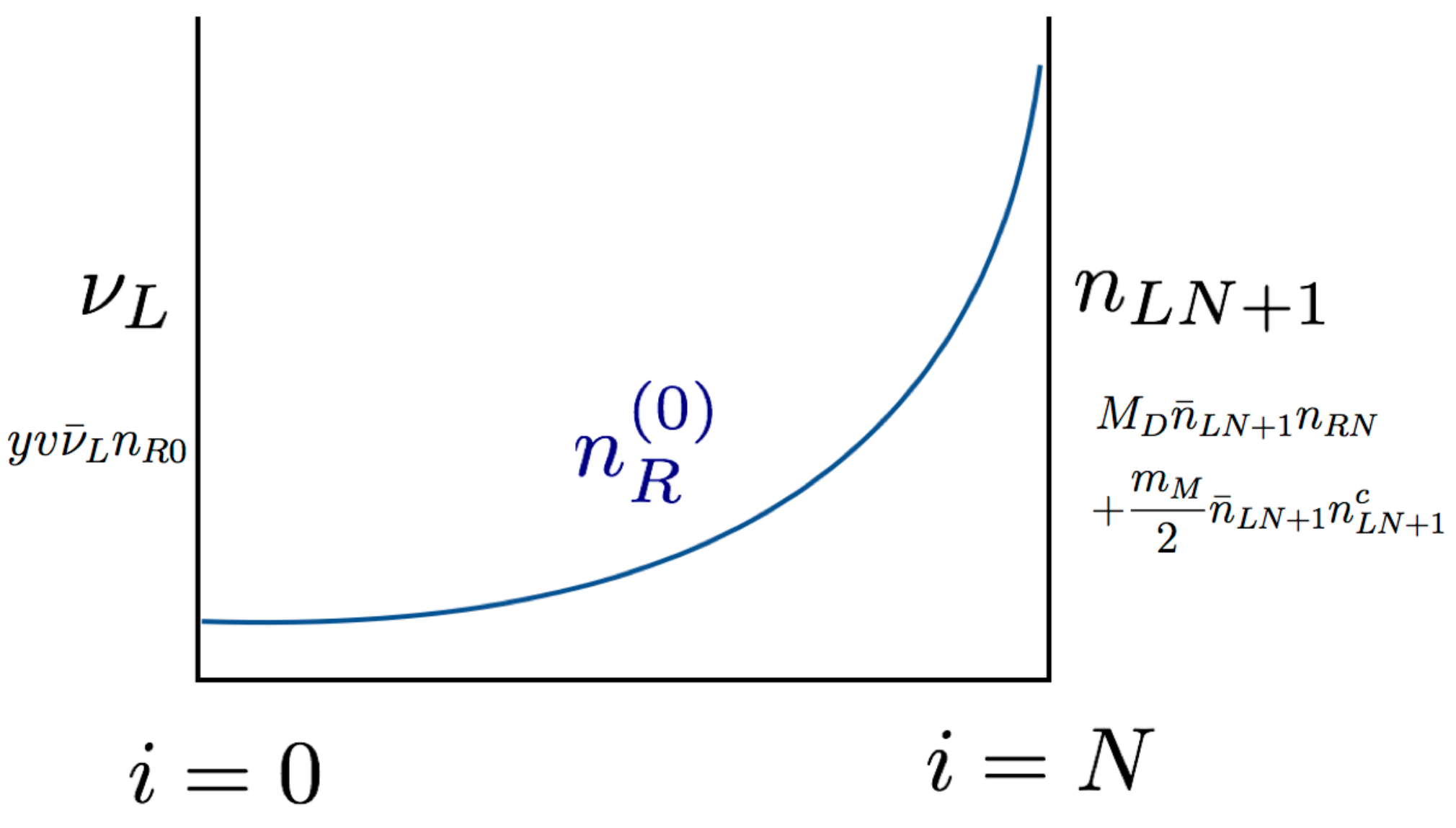}
\end{center}
\caption{Type Ia CW seesaw model.}\label{fig:inverse1}
\end{figure}

We first show how ``Inverse seesaw Case a" (Ia CW seesaw) is realized by a CW mechanism.  
In addition to the clockwork gears, CW$_R$, we introduce one more left-handed chiral fermion, $n_{L N+1}$. 
The potential is introduced as 
\begin{eqnarray}
V_{Ia} &=&  y v\bar\nu_L n_{R0}   + V_{{\rm CW}_{R,R}} +M_D\, \bar n_{L N+1} n_{R N}\nonumber\\
&&\, 
 + \frac{m_M}{2}  \bar n_{LN+1} n_{L N+1}^c + h.c.. 
\end{eqnarray}
Here we assume all mass parameters $M_D, m_M$ and $v$ have similar size.
The model is depicted in Fig.~\ref{fig:inverse1}.

It is noted that   in contrast to the type I CW seesaw model (Sec.~\ref{c_s}) ,
the total number of left-handed states, including the SM neutrinos, is greater than that of right-handed states. Therefore, 
without the Majorana mass term, $m_M$,  the SM neutrino remains massless. 
The easiest way of seeing the small neutrino mass  $m_\nu$ is to analyze the mass spectrum when $M_D,\, m_M < M$ where we may integrate out heavy modes of CW gears first.  The effective mass potential for 
$\nu_L,\, n_R^{(0)}$ and $n_{LN+1}$ is now given as 
\begin{eqnarray}
V_{Ia, \rm eff} &=&  y_{\rm eff} v \bar\nu_L n_R^{(0)} + M_D \bar n_{LN+1} n_R^{(0)}\nonumber\\
&&\,  +\frac{ m_M}{2} \bar n_{LN+1}  n_{LN+1} + h.c., 
\end{eqnarray}
with a suppressed effective Yukawa coupling, $y_{\rm eff} = y (m/M)^N$.
The  neutrino mass is read to be 
\begin{eqnarray}
m_\nu \simeq m_M\frac{y_{\rm eff}^2 v^2}{M_D^2}~~\text{(Type Ia)},
\end{eqnarray}
which indeed resembles the mass in inverse seesaw model (see Eq.~\ref{eq:mass IS}).
It is possible to get the full spectrum of the model by (numerically) diagonalizing the mass matrix {\it without} assuming
that the Dirac and Majorana mass terms are too small, in general, as it will be discussed more later. 

%%%%%%%%%%%%%%%%%%%%%%%%%%%%%%%%%%%%
%%%%%%%%%%%%%%%%%%%%%%%%%%%%%%%%%%%%

%%%%%%%%%%%%%%%%%%%%%%%%%%
\subsection{Type Ib CW  seesaw}
%%%%%%%%%%%%%%%%%%%%%%%%%%%

With the help of CW mechanism, small neutrino mass is realized when the CW zero mode has an exponentially small Majorana mass instead of suppressed Yukawa coupling, which corresponds to ``Inverse seesaw Case b" (Ib CW seesaw).   As a bonus of this realization, the unsuppressed Yukawa coupling may allow distinctive phenomenological features which can be tested by (future) experiments. 

\begin{figure}[t]
\begin{center}
\includegraphics[width=0.45\textwidth]{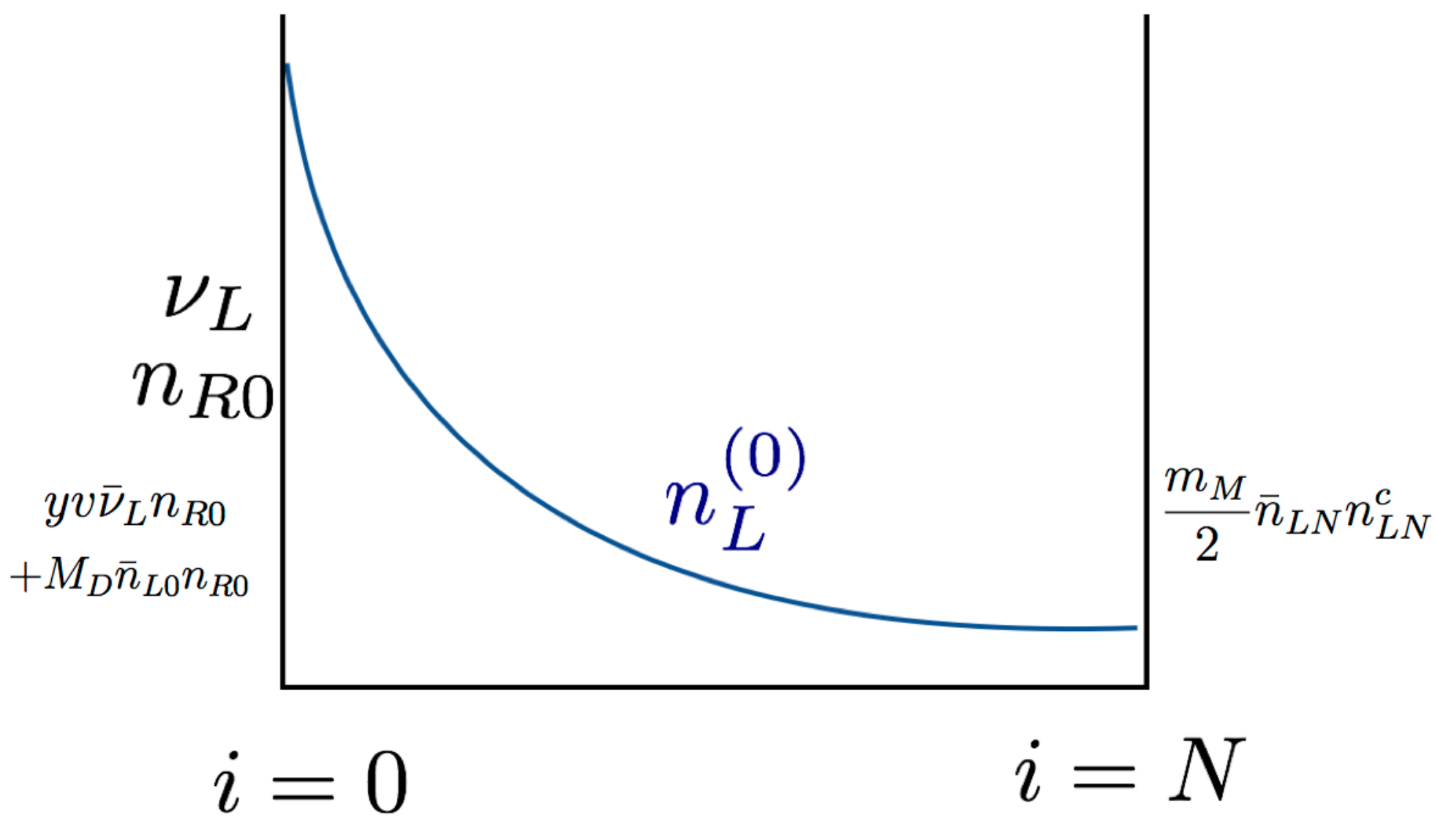} %figures directory 
\end{center}
\caption{Type Ib CW seesaw model.}\label{fig:inverse2}
\end{figure}

First, the left-clockwork gears in CW$_L$, are introduced to give the left-handed zero mode.  The clockwork potential is
\begin{eqnarray}
V_{{\rm CW}_{L, L}}= \sum_{i=1}^{N}   M \bar n_{Ri} n_{L i} - m\, \bar n_{Ri} n_{L\, i-1},
\end{eqnarray}
so that the zero mode is localized at $i=0$ site. 
We also introduce the right-handed singlet fermion, $n_{R0}$ and allow the mass terms
\begin{eqnarray}
V_{Ib} &=&  y v \bar\nu_L n_{R0}  + M_D \bar n_{L0} n_{R0}  + V_{{\rm CW}_{L,L}} \nonumber\\
&&\, + \frac{m_M}{2} \bar n_{L N} n_{L N}^c + h.c.,
\end{eqnarray} 
as the model is depicted in Fig.~\ref{fig:inverse2}. As it is seen in the figure, the zero mode profile is 
$n_{Li} \simeq (m/M)^i n_L^{(0)}$. 
The overlap between the Majorana mass terms and CW zero mode is very suppressed, and the effective Majorana mass is highly suppressed as  $m_M^{\rm eff} \equiv m_M (m/M)^{2N}$. 
The corresponding mass potential for the light fermions ($\nu_L,\, n_{R0}$, and $n_L^{(0)}$) is  obtained as
\begin{eqnarray}
V_{Ib, \rm eff} &=& y v \bar\nu_L n_{R0} + M_D \bar n_{L}^{(0)}n_{R0}\nonumber\\
&&\,  + \frac{m_M^{\rm eff}}{2}\bar n_L^{(0)} n_L^{(0)c} + h.c..
\end{eqnarray}
Finally, the neutrino mass is 
\begin{eqnarray}
m_\nu \simeq m_M^{\rm eff} \frac{y^2 v^2}{M_D^2  + y^2 v^2}~~\text{(Type Ib)},
\end{eqnarray}
as we desire.

%%%%%%%%%%%%%%%%%%%%%%%%%%%%%%%%%%%%
%%%%%%%%%%%%%%%%%%%%%%%%%%%%%%%%%%%%

\subsection{Extra dimensional interpretations}
\label{sec:xd}

The CW seesaw models in Fig.~\ref{fig:inverse1} and Fig.~\ref{fig:inverse2} have the extra dimensional interpretations:  the SM is localized on one boundary ($x_5=0$), whereas the lepton number violating operator  is localized on the opposite boundary $(x_5=L)$. The bulk fermion represents the clockwork gears whose zero mode profile is determined by the boundary conditions. In addition to the bulk fermion, there is the SM singlet fermion localized at $x_5=L$ for type Ia CW  and $x_5=0$ for type Ib CW inverse seesaw model, respectively. In both cases,  the quasi-localization of bulk fermion zero mode makes the couplings between the SM neutrino and the lepton number violating operator very suppressed, resulting in small neutrino masses~\cite{Park:2009cm}.

%%%%%%%%%%%%%%%%%%%%%%%%%%%
\section{Discussion}
\label{sec:discussion}
%%%%%%%%%%%%%%%%%%%%%%%%%%%

We have  classified three types of CW seesaw models with and without Majorana mass terms,   especially we have newly proposed the clockwork realization of the inverse seesaw mechanism.
The details of realization would differ among the realization but they have common features: no tuning in the model but successful generation of small neutrino mass. 

Phenomenological distinction would be possible by looking into full spectrum of each realization:
In each realization, besides the zero mode whose couplings are hierarchically suppressed, 
there are heavy modes which have the sizable couplings to matters in general in CW models.
Those heavy modes may be within the reach of upcoming experiments thus should be clarified. 

Basically, there are $2N$ number of heavy Majorana (approximately Dirac) fermions of clockwork gears with masses of ${\cal O}(M\pm m)$ \cite{Giudice:2016yja}. In addition to those modes, the type I model has a relatively light fermion:
\begin{eqnarray}
&&\textrm{Type I}:\, M_{\rm I}  = m_M, 
\end{eqnarray}
for $m_M < M$. Type Ia,b models show different possibilities. For $M_D \sim v  < M$,  
there are two relatively light Fermis with masses, 
\begin{eqnarray}
&&\textrm{Type Ia}:\, M_{\rm Ia\pm}  = M_D \pm \frac{m_M}{2}  + \frac{m_M^2}{8 M_D}\pm \frac{y_{\rm eff}^4 v^4m_M^3}{2M_D^6},  \nonumber\\
&&\textrm{Type Ib}:\, M_{\rm Ib\pm} =  \sqrt{M_D^2 + \lambda^2 v^2} 
\pm \frac{ m_M^{\rm eff} M_D^2}{2(M_D^2 +\lambda^2 v^2)}. \nonumber \\
\end{eqnarray}
Theses massive fermions are essentially decoupled in Type I and Type Ia models because of small Yukawa couplings, $y_{\rm eff}$. However in Type Ib model, they have the sizable couplings so that we may directly observe those modes e.g. at the collider experiments (see e.g.~\cite{Park:2009cm,Agashe:2016ttz, Agashe:2017ann} and ~\cite{park22}). 

Let us comment on the full mass matrix of the fermions in Type I inverse seesaw models for $M_D\sim M$, as a result,  we also have to consider full CW gears simultaneously.  It is noted that  type Ia and Ib  share the same mass matrix  if we take $M_D= M$ and match $N_{Li+1}$ in Type Ia to $N_{Li}$ in Type Ib. In this simplest example, 
for $\Psi = (\nu_L,\, n_{R0},\, n_{L0},\,\cdots, n_{RN},\, n_{LN})^T$,   
the potential has the form 
$ V=  \frac{1}{2} \bar\Psi {\cal M}\Psi^c + h.c.$, where \begin{eqnarray} \label{massmatrix}
{\cal M} = \left(\begin{array}{ccccccc} 
   0 & y v & 0 & 0  &  \cdots  &  0 & 0 \\
y v &  0   & M & 0  &  \cdots  & 0 & 0 \\
  0  & M  &  0  & -m & \cdots  & 0 & 0\\
  0 &  0  & -m  & 0   &  \cdots & 0 & 0\\
    \vdots  & \vdots  &  \vdots & \vdots & \ddots & \vdots  & \vdots \\
  0 &  0  & 0  & 0   &    \cdots & 0 & M\\
  0 & 0   & 0   & 0   &   \cdots    &  M & m_M\end{array} \right).
\end{eqnarray}
By diagonalizing the mass matrix, we get  the full spectrum of light and heavy modes. 

The mass of the lightest fermion  is
\begin{eqnarray}
m_\nu 
\approx m_M \left(\frac{y v}{M}\right)^2\left(\frac{m}{M}\right)^{2N} \left(\frac{M^2-m^2}{M^2 +y^2v^2 - m^2}\right). 
\end{eqnarray}
This result shows that the essential feature of our set-up that $m_\nu$ is proportional to  $(m/M)^{2N}$ as the result of the CW zero mode localization is rather insensitive to the size of Dirac and Majorana mass terms for   boundary gear fields.   We could understand the factor $(m/M)^{2N}$ as following.   
In Eq.~(\ref{massmatrix}), we relabel $m$ as $m_i$ from $i=1,\cdots, N$ for 
the $(2i+1)\times (2i+2)$, and $(2i+2)\times (2i+1)$ components.  If we turn off the $m_i$, then 
Eq.~(\ref{massmatrix}) is divided by two block diagonal  
$(2i+1)\times (2i+1)$, and $2(N-i+1)\times 2(N-i+1)$ matrices. 
The neutrino mass is coming from $(2i+1)\times (2i+1)$ matrix, which becomes zero. 
From this argument, we get $m_\nu\propto \prod_{i=1}^N m_i$. By dimensional analysis, 
$m_\nu \propto m_N \prod_{i=1}^N (m_i/M)= m_N (m/M)^N$. This implies that the CW mechanism for the neutrino mass is  robust from boundary interactions, and we can easily identify it numerically. 
For example, in the case with type Ia,  we could take $N=3$, $y=1$, $M=10 v$,  $M_D=100 v$, $m=v$, and $m_M= 2v$.
The resulting neutrino mass becomes $m_\nu = 1.98\times 10^{-10}v$ numerically. 
This is well matched with the analytic approximation, $m_\nu = m_M(yv/M_D)^2(m/M)^{2N}$.  
For the heavy fermions, the couplings in $c_{(n)} \bar l_L H n^{(n)}_R $
are  of the form of $c_{(n)} ={\cal O}(1/\sqrt{N}- 1/\sqrt{N^3})$ and the masses are
$M_{(n)} = M + {\cal O}(m,\, yv,\, m_M)$. The heavy modes would decay as $n_R^{(n)} \to h \nu_L$ with (un)suppressed coupling strengths. 

Finally, the unsuppressed  interactions have interesting cosmological implication as heavy modes and the SM Higgs boson and the neutrino are in touch and $n_R^{(n)} \leftrightarrow h \nu_L$ conversion takes place then decouples successively at  $T_{(n)}\sim M_{(n)}$. Since they are the source of lepton number violating interactions, there could be the important predictions for the mass spectrum from the effect on baryo-leptogenesis~\cite{park22}.

\vspace{.4cm}

%%%%%%%%%%%%%%%%%%%%%%%%%%%%%%%%%%%%%%%%%%%%%%%%
%%%%%%%%%%%%%%%%%%%%%%%%%%%%%%%%%%%%%%%%%%%%%%%%
\acknowledgments
This work was initiated at the CERN-CKC workshop
``What's going on at the weak scale?'' in Jeju Island, Korea in June 2017.
SCP is supported by the National Research Foundation of Korea (NRF) grant funded by the Korean government (MSIP) 
(No. 2016R1A2B2016112). CSS acknowledges  the support from the Korea Ministry of Education, Science and Technology, Gyeongsangbuk-Do and Pohang City for Independent Junior Research Groups at the Asia Pacific Center for Theoretical Physics. CSS is also supported by the Basic Science Research Program through the NRF Grant (No. 2017R1D1A1B04032316).
 
%%%%%%%%%%%%%%%%%%%%%%%%%%%%%%%%%%%%%%%%%%%%%%%%
%%%%%%%%%%%%%%%%%%%%%%%%%%%%%%%%%%%%%%%%%%%%%%%%

\bibliographystyle{apsrev4-1}
\bibliography{cw}

\end{document}